\documentclass[pra,showkeys,showpacs,notitlepage,twocolumn]{revtex4-2}

\usepackage{dcolumn}
\usepackage{amsmath}

\usepackage{graphicx}
\usepackage{fancyvrb}

\usepackage{color}
\definecolor{brown}{rgb}{0.8,0.6,0.3}
\definecolor{dgreen}{rgb}{0.2,0.4,0.3}

\usepackage[pdfauthor={Richard J. Mathar}, colorlinks=true,urlcolor=blue,linkcolor=red,citecolor=brown,
pdfkeywords={Kepler Equation, Inverse Problem},
pdftitle={First Estimates of Kepler's Equation},
bookmarksopen=true,pdfpagelayout=OneColumn]{hyperref}

\newcommand{\sgn}[1]{\mathrm{sgn}(#1)}
\DeclareMathOperator{\arsinh}{arsinh}

\begin{document}

\title{Improved First Estimates to the Solution of Kepler's Equation}

\author{Richard J. Mathar}
\homepage{https://www.mpia-hd.mpg.de/~mathar} 
\affiliation{Max-Planck Institute of Astronomy, K\"onigstuhl 17, 69117 Heidelberg, Germany} 

\date{\today}

\begin{abstract}
The manuscripts provides a novel starting guess for the solution of Kepler's equation 
for unknown eccentric anomaly $E$ given the eccentricity $e$ and mean 
anomaly $M$ of an elliptical orbit. 
\end{abstract}

\maketitle

\section{Kepler's Equation}
\subsection{Mean and Eccentric Anomaly}

The track of the orbit for a 2-body potential proportional to the
inverse distance of the two bodies leads to solutions which may be
ellipses with eccentricity $0\le e\le 1$. The time dependence
is described by the parameter $M$ of the mean anomaly, which 
is an angle measured from the center of the ellipse,
and which is a product of a parameter $n$ called the mean motion
(essentially the square root of the coupling parameter
in the numerator of the 2-body potential divided by the cube
of the major semi-axis)
and a time elapsed since some reference epoch $t_0$:
\begin{equation}
M=n(t-t_0).
\end{equation}

For the manuscript at hand, $M$ and $e$ are considered fixed parameters.
To compute the circular coordinates of distance and
true anomaly of the body at that time in
the reference frame centered at the ellipse, one encounters
Kepler's equation
\begin{eqnarray}
E&=&M+e\sin E \label{eq.E}\\
E-M-e\sin E&=&0. \label{eq.f}
\end{eqnarray}
$E$ and $M$ are angles measured in radian in the range $-\pi \le E,M\le \pi$.
To simplify the notation, we discuss only the cases where $M\ge 0$, because the parity
\begin{equation}
E(-M)=-E(M),
\end{equation}
---equivalent to flipping the entire orbit along the major axis
of the ellipse---allows to recover solutions for negative $M$ as well.

\subsection{The Inverse Problem}

The numerical problem considered here is to find the root of
the function
\begin{equation}
f(E) \equiv E-M-e\sin E
\end{equation}
in an efficient and numerically stable fashion.

The expansion of $E$ in a Taylor series of $e$
can be written as
\begin{equation}
E = \sum_{i\ge 0} m_i(M) e^i
\label{eq.Eofe}
\end{equation}
supported by the table \cite[(4.2.3)]{FitzpatrickPM}

\begin{tabular}{l|l}
$i$ & $m_i(M)$ \\
\hline
0 & $M$ \\
1 & $S$ \\
2 & $CS$ \\
3 & $S(3C^2-1)/2$\\
4 & $CS(8C^2-5)/3$\\
5 & $S(125C^4-114C^2+13)/24$\\
6 & $CS(162C^4-194C^2+47)/15$\\
7 & $S(16087C^6-24915C^4+9369C^2-541)/720$\\
8 & $CS(16384C^6-28950C^4+14838C^2-1957)/325$\\
\end{tabular}

where  we have written $C\equiv \cos M$ and $S\equiv \sin M$
to tighten the notation.
The standard problem with this series is that the $m_i$ do not
fall in magnitude as a function of $i$; so the Taylor expansion is
not converging well unless $e$ is close to zero.

\section{Newton Methods}
The simplest technique of solving (\ref{eq.E}) is a fixed point iteration
\begin{equation}
E^{(i+1)}=M+e\sin E^{(i)};\quad E^{(0)}=M. 
\label{eq.fpt}
\end{equation}
For faster convergence this is commonly replaced by a first-order Newton iteration
\begin{equation}
E^{(i+1)}= E^{(i)}-\frac{f}{f'},\quad f'\equiv \partial f/\partial E,
\label{eq.N1}
\end{equation}
or a second-order Newton iteration
\cite{DanbyCM31,GerlachSIAM36,HansenNumerMath27,KalantariJCAM80,AlefeldAMM88,EsmaelIJCA89}
\begin{equation}
E^{(i+1)}=E^{(i)} -\frac{f}{f'(1-\frac{ff''}{2f^{'2}})}
= E^{(i)}-\frac{2ff'}{2f^{'2}-ff''},
\label{eq.N2}
\end{equation}
where the function and its derivatives with respect to the unknown $E$ are
\begin{equation}
f \equiv E-e\sin E-M;\quad
f' \equiv 1-e\cos E;\quad
f'' \equiv e\sin E.
\end{equation}
(\ref{eq.N1}) is for example used in the \texttt{iauPlan94} IAU function for
planets' ephemerides \cite{SOFA}.
Note that, since the evaluation of the trigonometric functions is expensive
compared to the fundamental operations \cite{FukushimaCM66,PalaciosJCAM138}, the second-order
iteration is preferred since $\sin E$ in $f''$ is already 
calculated in conjunction with $f$.

Higher-order algorithms are applicable \cite{GrauAMC218,ChunAMC190}.

\section{Initial Value Problem}

\subsection{Standard Initial Guesses}

\begin{figure}[htp]
\includegraphics[scale=0.5]{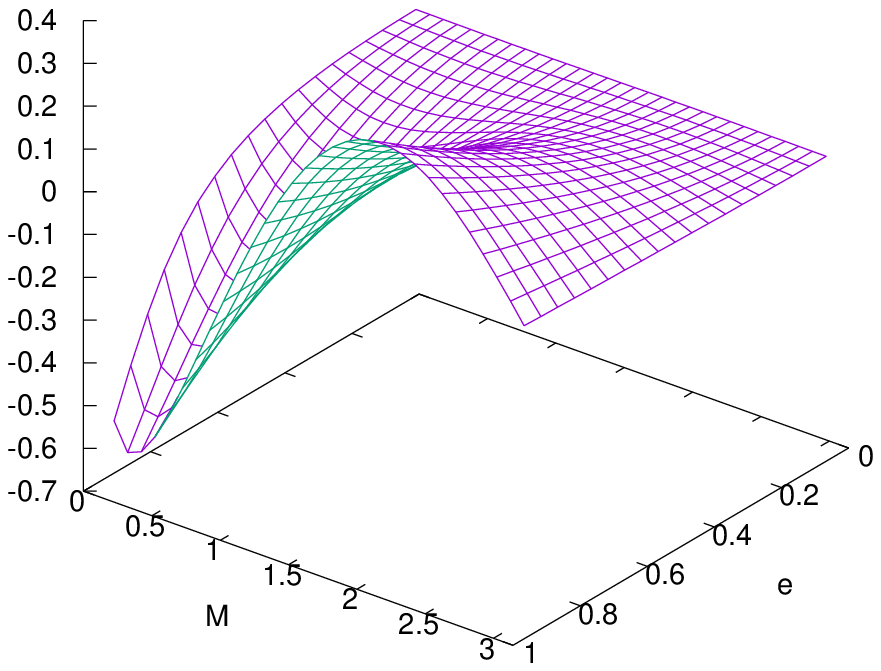}
\caption{Mismatch $E^{(0)}-E$ of the initial estimate (\ref{eq.E0fp}).}
\label{fig.plotFp2}
\end{figure}

If the initial guess is the second step of (\ref{eq.fpt}), 
\begin{equation}
E^{(0)} =M+e\sin M,
\label{eq.E0fp}
\end{equation}
and the 
iteration (\ref{eq.N1}) is used with $e>0.99$,
a known problem is that the iterations may converge to secondary roots
of the equation with the wrong sign \cite{ConwayAIAA86}.
This is basically triggered by starting with an underestimate of $E$
as illustrated in Figure \ref{fig.plotFp2}.

\begin{figure}[htp]
\includegraphics[scale=0.5]{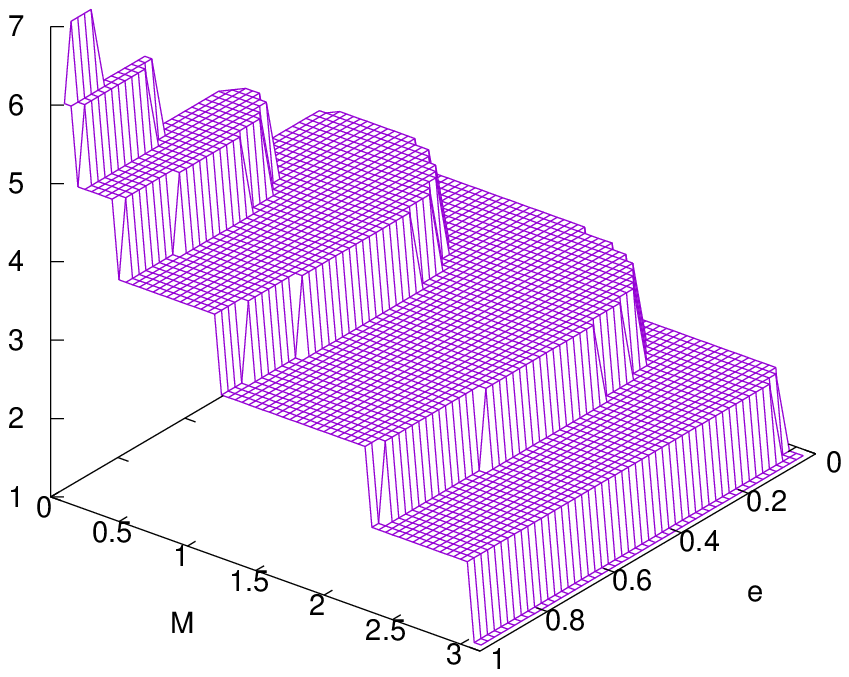}
\caption{
The number of iterations needed for a relative accuracy of $10^{-12}$ in $E$
starting from (\ref{eq.E0Pi}) iterating with (\ref{eq.N1}).
}
\label{fig.dgsPi1}
\end{figure}

\begin{figure}[htp]
\includegraphics[scale=0.5]{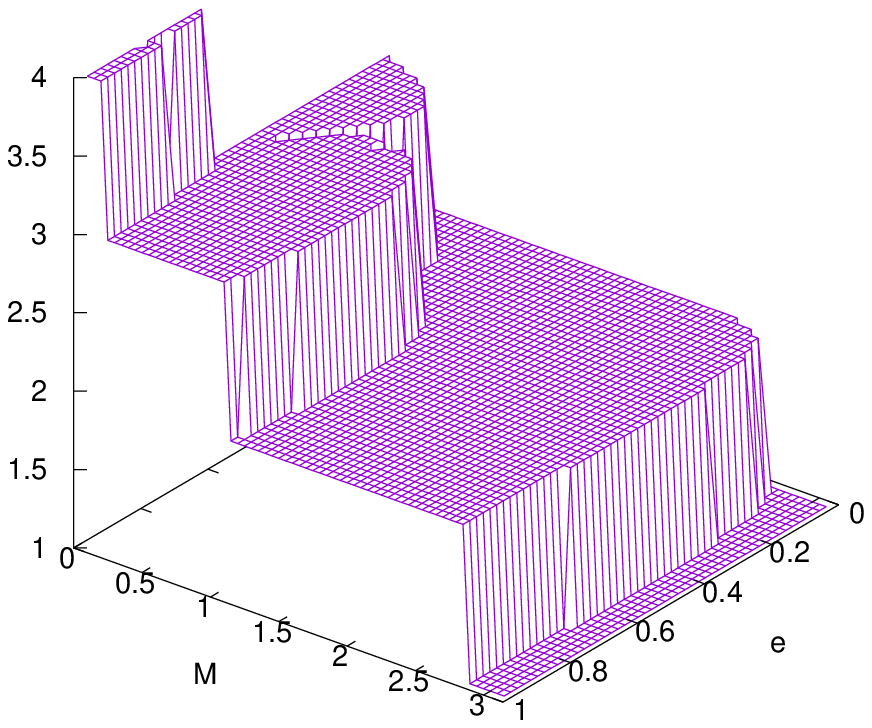}
\caption{
The number of iterations needed for a relative accuracy of $10^{-12}$ in $E$
starting from (\ref{eq.E0Pi}) iterating with (\ref{eq.N2}).
}
\label{fig.dgsPi2}
\end{figure}

A well-known remedy is to start with the initial guess
\begin{equation}
E^{(0)}=\pi
\label{eq.E0Pi}
\end{equation}
which is known to converge \cite{CharlesCMDA69,StumpfCM74}. The
speed of convergence with the two basic Newton methods is illustrated
in Figures \ref{fig.dgsPi1} and \ref{fig.dgsPi2}.
\subsection{Eccentricity One}
If $e=1$, $M=E-\sin E$ has the power series $M = \sum_{i\ge 1} (-)^{i-1}E^{2i+1}/(2i+1)!$.
The associate power series for the cube root
\begin{equation}
\sqrt[3]{6M}\equiv \bar M =E-\frac{1}{60}E^3+\frac{1}{8400}E^5+O(E^9),
\end{equation}
can be reversed \cite[(3.6.25)]{AS}
\begin{multline}
E=\bar M +\frac{1}{60}\bar M^3+\frac{1}{1400}\bar M^5+\frac{1}{25200}\bar M^7\\
+\frac{43}{17248000}\bar M^9+\frac{1213}{7207200000}\bar M^{11}\\
+\frac{151439}{12713500800000}\bar M^{13}
+\frac{33227}{38118080000000}\bar M^{15}\\
+\frac{16542537833}{252957982717440000000}\bar M^{17}
+\cdots
\label{eq.Mcub}
\end{multline}
Again this is not converging well to $\to \pi$ as $M\to \pi$,
but since $E$ is an increasing function of $e$ at constant $M$,
this approximation is slightly better than \eqref{eq.E0Pi} as an
initial estimator from above.
\begin{figure}[htp]
\includegraphics[scale=0.5]{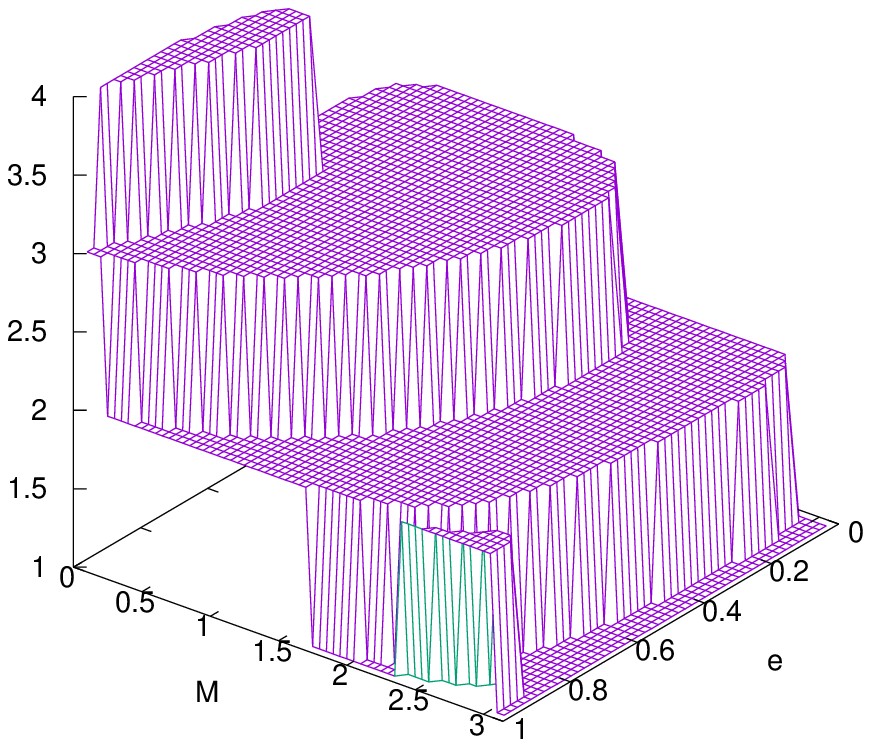}
\caption{
The number of iterations needed for a relative accuracy of $10^{-12}$ in $E$
starting from \eqref{eq.Mcub}, terms up to including $O(\bar M^7)$, 
iterating with (\ref{eq.N1}).
}
\label{fig.dgsMcub1}
\end{figure}
This is illustrated by the lower number of iterations in
Fig.\ \ref{fig.dgsMcub1} compared to Fig. \ref{fig.dgsPi1}.

\subsection{Taylor series at various $M$}
The derivate of $E$ with respect to $M$ at constant $e$ is according to \eqref{eq.E}
\begin{equation}
\frac{dE}{dM} = 1 +e\cos E \frac{dE}{dM},
\end{equation}
or solved for $dE/dM$
\begin{equation}
dE/dM =\frac{1}{1-e\cos E}.
\end{equation}
Repeated differentiation and using the initial value $M=E=\pi$
builds a Taylor expansion of $E$ in powers of $M-\pi$:

\begin{multline}
E=\pi +\frac{1}{1+e}(M-\pi)+\frac{e}{(1+e)^4}\frac{(M-\pi)^3}{3!}\\
+\frac{e(9e-1)}{(1+e)^7}\frac{(M-\pi)^5}{5!}\\
+\frac{e(1-54e+225e^2)}{(1+e)^{10}}\frac{(M-\pi)^7}{7!}
+\cdots
\label{eq.TaylPi}
\end{multline}
This approximation as the starting value 
has excellent quality for $M>1$.
With the same method the Taylor expansion with the initial
value $E=M=0$ is constructed:
\begin{multline}
E=\frac{1}{1-e}M+\frac{e}{(1-e^2)^2}\frac{M^3}{3!}\\
-\frac{e(1-8e-e^2)}{(1-e^2)^4}\frac{M^5}{5!}\\
+\frac{e(1-52e+170e^2+52e^3-35e^4)}{(1-e^2)^6}\frac{M^7}{7!}
+\cdots,
\label{eq.Tayl0}
\end{multline}
but this is only advantageous if $e<0.5$.
A third variant is to build a Taylor expansion around $M=\pi/2-e$, $E=\pi/2$:
\begin{multline}
E=\frac{\pi}{2}
+(M-\frac{\pi}{2}+e)
+\frac{e}{(1+e)^2}\frac{(M-\frac{\pi}{2}+e)^3}{3!}\\
-\frac{e(1-8e-5e^2)}{(1+e)^4}\frac{(M-\frac{\pi}{2}+e)^5}{5!}\\
+\frac{e(1-52e+144e^2+224e^3+61e^4)}{(1+e)^6}\frac{(M-\frac{\pi}{2}+e)^7}{7!}
+\cdots,
\label{eq.TaylPih}
\end{multline}
A fourth variant is to build a Taylor expansion around $M=\pi/6-e/2$, $E=\pi/6$:
\begin{multline}
E=\frac{\pi}{6}
+(M-\frac{\pi}{6}+\frac{e}{2})\frac{1}{1-\sqrt{3}e/2}\\
+\frac{e}{(1+e)^2(1-\sqrt{3}e/2)^2}\frac{(M-\frac{\pi}{6}+\frac{e}{2})^3}{3!}\\
-\frac{(2\sqrt{3}-5)e(-2\sqrt{3}-5+16\sqrt{3}e+40e+13e^2)}{13(1+e)^4(1-\sqrt{3}e/2)^4}\frac{(M-\frac{\pi}{6}+\frac{e}{2})^5}{5!}\\
+\cdots,
\label{eq.TaylPi6}
\end{multline}
The relative merits of these 4 Taylor series are a complicated function of $e$ and $M$.
As a guideline
\begin{itemize}
\item
for simplicity \eqref{eq.TaylPih} would not be used at all;
\item
\eqref{eq.TaylPi} be used where $\frac34 (1-e)<M\le \pi$;
\item
\eqref{eq.Tayl0} be used where $0\le M<\frac14-\frac12 e$;
\item
\eqref{eq.TaylPi6} be used elsewhere.
\end{itemize}

\cleardoublepage
\subsection{Improved Initial Value, Version 1}
A starting value of $E$ is obtained by inserting the approximation
\begin{equation}
\sin E\approx 1-\frac{4}{\pi^2}(E-\pi/2)^2
\end{equation}
into the equation.
[Similar approximations could be obtained by truncating the Chebyshev
series approximation of the $\sin E$ after the second term \cite{Schonfelder,BoydANM57,BoydJCAM223}.]

\begin{figure}[htp]
\includegraphics[scale=0.5]{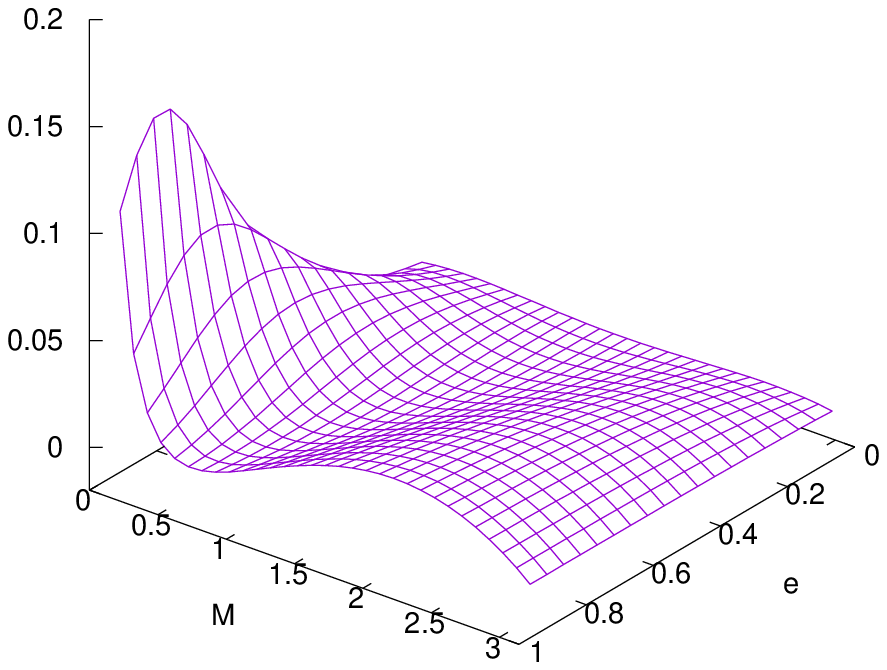}
\caption{Mismatch $E^{(0)}-E$ of the initial estimate (\ref{eq.E0qu}).}
\label{fig.plotE2}
\end{figure}

\begin{figure}[htp]
\includegraphics[scale=0.5]{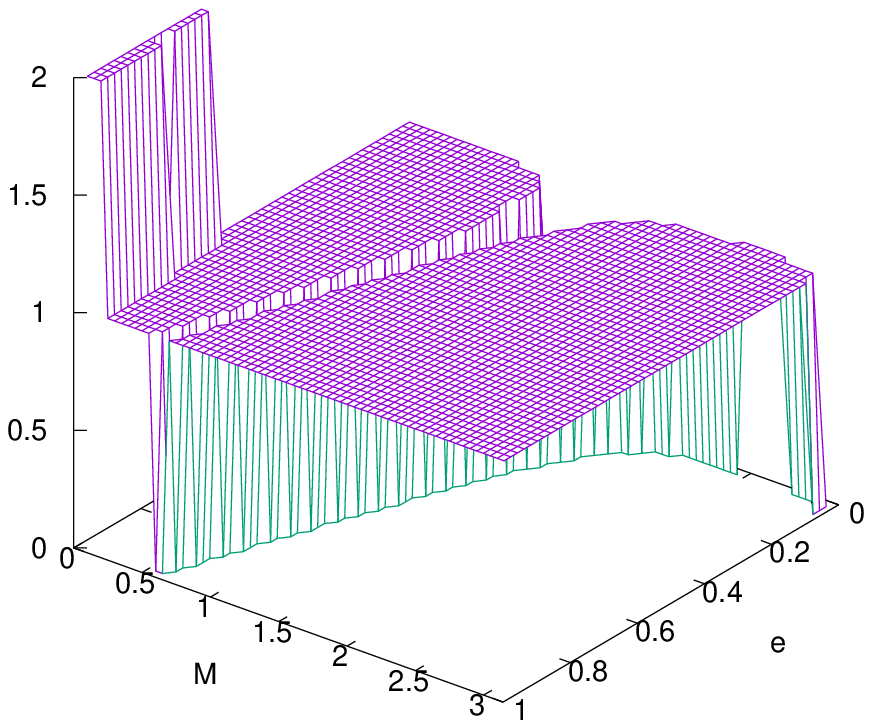}
\caption{
The number of iterations needed for a relative accuracy of $10^{-12}$ in $E$
starting from (\ref{eq.E0qu}) iterating with (\ref{eq.N2}).
}
\label{fig.dgsEq2}
\end{figure}

This leads to a quadratic equation for $E$ 
\begin{equation}
E^{(0)}=M+e\left[1-\frac{4}{\pi^2}(E^{(0)}-\pi/2)^2\right],
\end{equation}
which is solved by
\begin{equation}
\bar e\equiv \frac{\pi}{4e}-1 ;
\end{equation}
\begin{equation}
E^{(0)}
=\frac{\pi}{2}\bar e
\left[\sgn{\bar e} \sqrt{1+\frac{M}{e {\bar e}^2}}-1\right].
\label{eq.E0qu}
\end{equation}
The error of this estimate relative to the accurate solution
is shown in Figure \ref{fig.plotE2}. It increases where $M/e\to 0$
and $\bar e\to 0$.
The figure shows that the
(\ref{eq.E0qu}) has the same benefit as (\ref{eq.E0Pi}) of approximating the
solution from above, therefore converging \cite{CharlesCMDA69}, but being more accurate.
In consequence, the convergence is faster, as demonstrated
in Figure  \ref{fig.dgsEq2} if compared with Figure \ref{fig.dgsPi2}.

\cleardoublepage

\subsection{Improved Inital Value, Version 2}

If (\ref{eq.f}) is expressed as
\begin{equation}
E-M=e\sin(M+E-M),
\end{equation}
both sides may be expanded in a Taylor series of $E-M$,
\begin{equation}
E-M\approx e\sin(M)+(E-M)e\cos M -\frac{(E-M)^2}{2}e\sin M+\ldots
\label{eq.emdiff}
\end{equation}
Keeping this series up to $O(E-M)$ yields the estimate
\begin{equation}
E^{(0)}=M+\frac{e\sin(M)}{1-e\cos M}.
\label{eq.E0t1}
\end{equation}
This is basically the estimate of the second step of the
fixed point iteration (\ref{eq.fpt}) with an 
enhancement factor of the second term if $e$ or $\cos M$ are large.
As pointed out earlier \cite{DanbyCM31}, this is also obtained applying the Newton
method to the estimator $E^{(0)}=M$.
\begin{figure}[htp]
\includegraphics[scale=0.5]{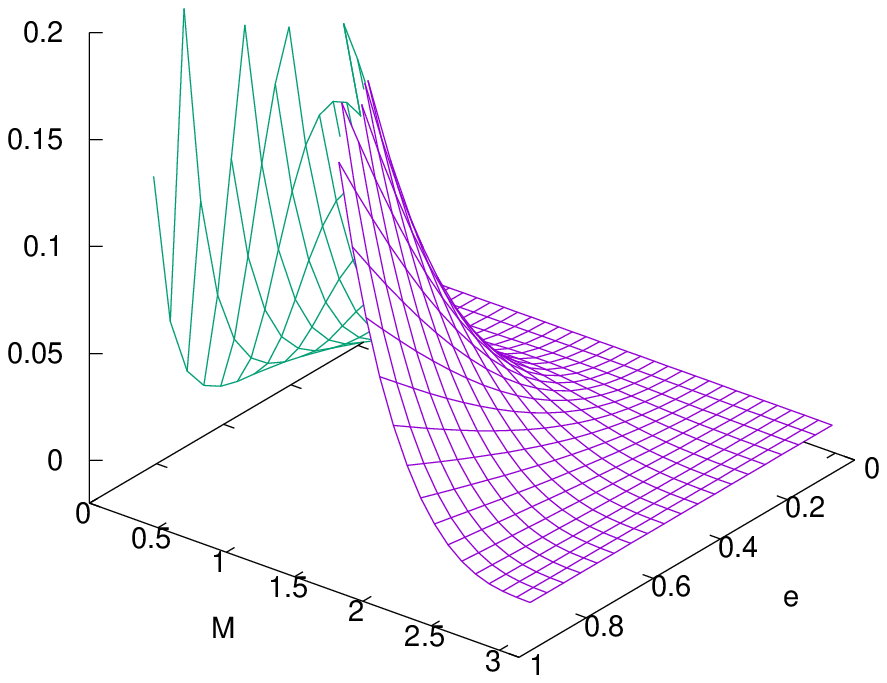}
\caption{Mismatch $E^{(0)}-E$ of the initial estimate (\ref{eq.E0t1}).}
\label{fig.E0T1}
\end{figure}

If the series is kept up to $O((E-M)^2)$, the associated quadratic
equation proposes
\begin{equation}
\frac{e}{2}\sin M (E^{(0)}-M)^2+(1-e\cos M)(E^{(0)}-M)-e\sin M=0.
\end{equation}
This quadratic equation is solved by
\begin{equation}
E^{(0)}-M=
\frac{1-e\cos M}{e\sin M}
\left[
\sqrt{1+\frac{2e^2\sin^2M}{(1-e\cos M)^2}}
-1
\right]
.
\label{eq.E0t2}
\end{equation}
\begin{figure}[htp]
\includegraphics[scale=0.5]{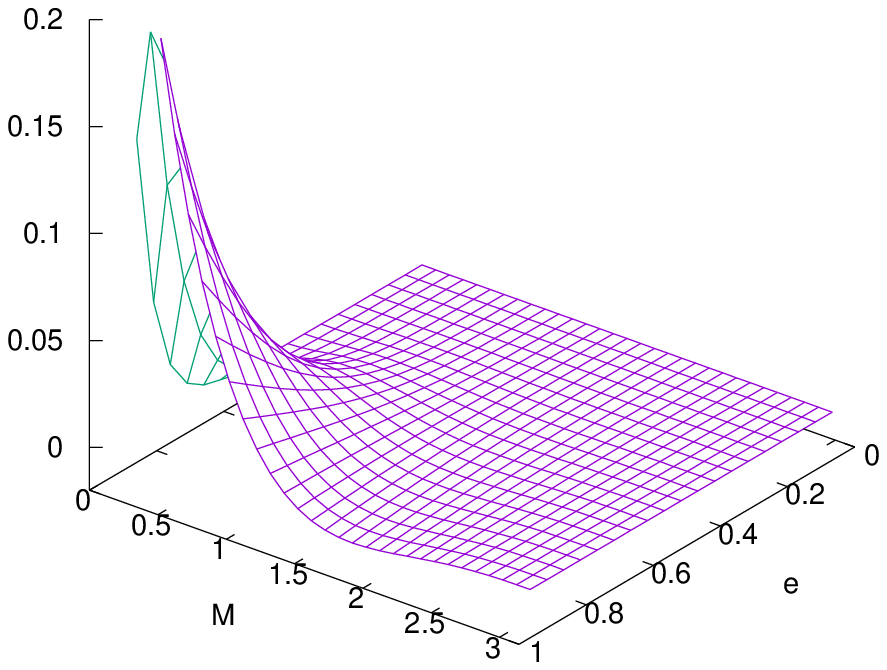}
\caption{Mismatch $E^{(0)}-E$ of the initial estimate (\ref{eq.E0t2}).}
\label{fig.E0T2}
\end{figure}
Figures \ref{fig.E0T1} and \ref{fig.E0T2} show in comparison
with Figure \ref{fig.plotE2} that
these approximations derived from the Taylor series of $E-M$
are not better than the one from the quadratic estimate of $\sin E$.

\begin{figure}[htp]
\includegraphics[scale=0.5]{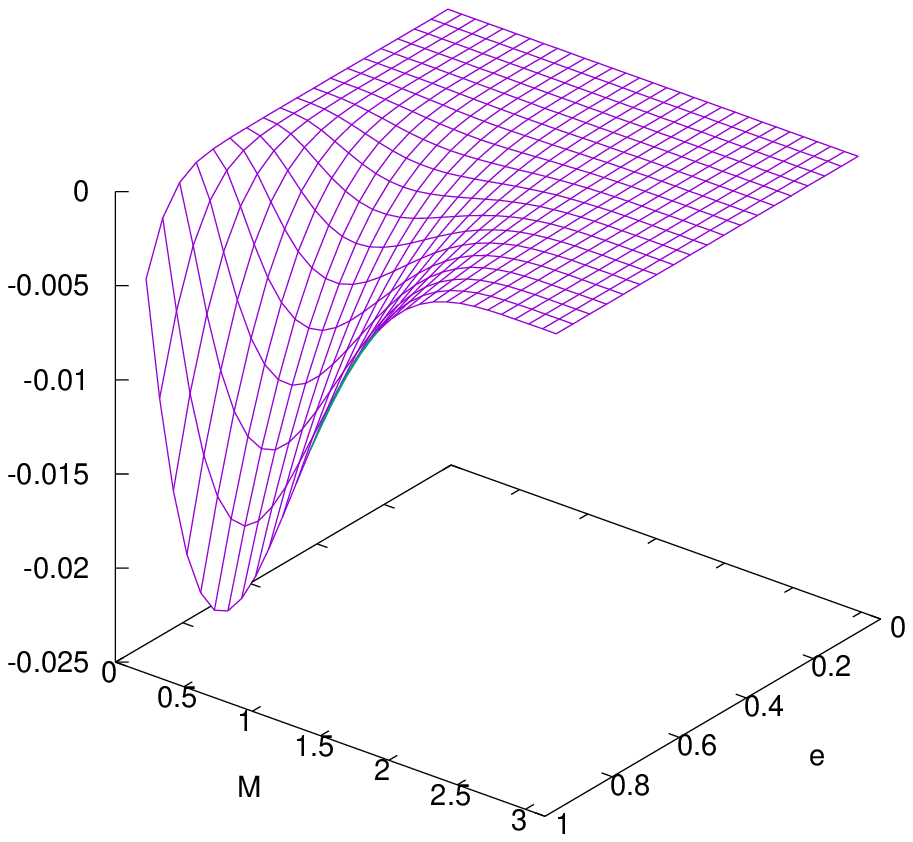}
\caption{Mismatch $E^{(0)}-E$ of the initial estimate of the cubic order of (\ref{eq.emdiff}).}
\label{fig.E0T3}
\end{figure}

Expansion of (\ref{eq.emdiff}) up to third order in $E-M$ yields a cubic equation
for $E^{(0)}-M$, which is even closer to the exact solution
as demonstrated in Figure \ref{fig.E0T3}.
See Mikkola's paper for a similar approach \cite{MikkolaCM40,NijenhuisCM51,MarkleyCM63}.

In a systematic treatment,
\eqref{eq.emdiff} is power series of $e\sin M$ in powers of $E-M$,
\begin{equation}
eS=
(1-eC)(E-M)+\frac{eS}{2}(E-M)^2+\frac{eC}{6}(E-M)^3+\cdots
\end{equation}
with series inversion \cite[3.6.25]{AS}
\begin{multline}
E-M = 
\frac{1}{1-eC}(eS)
-\frac{eS}{2(1-eC)^3}(eS)^2\\
+\frac{e(e+2eS^2-C)}{6(1-eC)^5}(eS)^3\\
-\frac{eS(-8eC-1+9e^2+6e^2S^2)}{24(1-eC)^7}(eS)^4+\cdots,
\end{multline}
but this right hand side does not converge well for $eC>0.5$.

\subsection{Adiabatic switching on}
\begin{figure}[htp]
\includegraphics[scale=0.5]{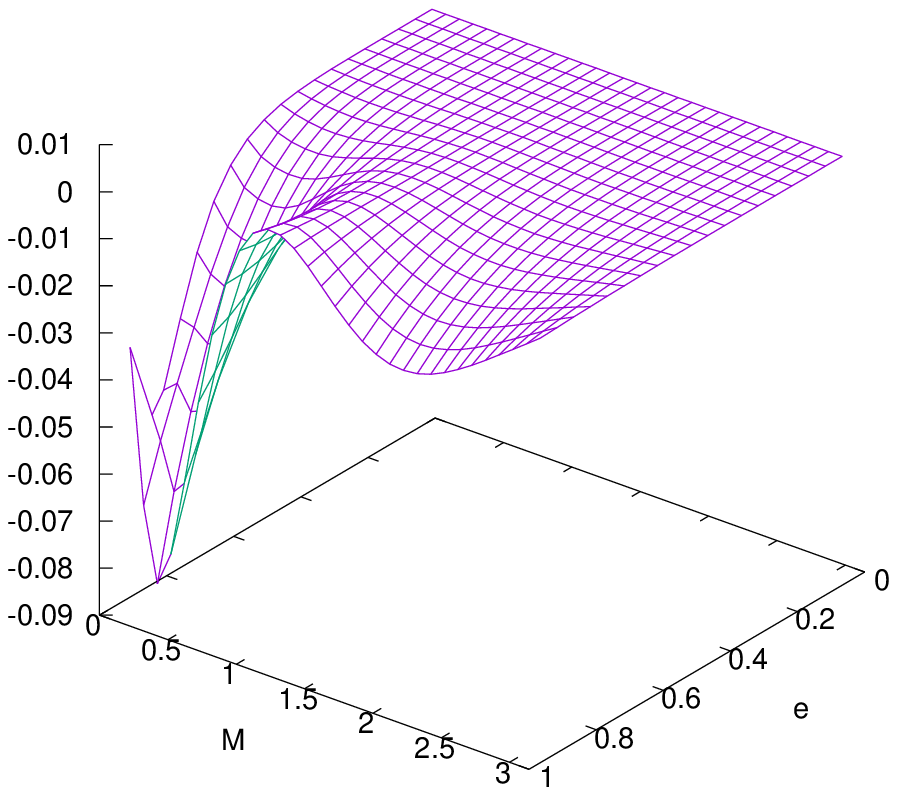}
\caption{Mismatch $E^{(0)}-E$ of the initial estimate of a single-step
Runge-Kutta estimator for \eqref{eq.RK4}.
}
\label{fig.RK4}
\end{figure}

One principle in perturbative quantum mechanics 
switches on the fermionic interaction by increasing 
the coupling (fine structure) constant from zero (no interaction) up to the
value attained by the real-world system. Adopting this
concept here, the eccentricity is started
at $e=0$ at the known solution $E=M$, and the $E$ is tracked
until the actual value of $e$ is reached. Let overdots denote
partial derivatives with respect to $e$ at constant $M$, e.g. $\dot e=1$, $\dot M=0$. 
The derivative of \eqref{eq.E} is
\begin{equation}
\dot E = \sin E+e\cos E \dot E ;
\end{equation}
\begin{equation}
(1-e\cos E)\dot E = \sin E.
\label{eq.dotE}
\end{equation}
To avoid numerically expensive evaluations of the trigonometric functions
the auxiliary angle $\phi\equiv \cos E$ is introduced with derivative
\begin{equation}
\dot\phi = -\sin E \dot E.
\end{equation}
Multiplying \eqref{eq.dotE} with $\dot E$ yields a nonlinear first order
differential equation for $\phi$,
\begin{equation}
\dot \phi = -\frac{1-\phi^2}{1-e\phi}
.
\label{eq.RK4}
\end{equation}
We solve this with a single-step ``explicit'' classic Runge-Kutta
integration with the initial value $\phi(e=0)=\cos M$ up to the
actual $e$ \cite{ZurmuhlZAMM28,ButcherJAMS3}.  These estimates $E$ are surprisingly close to the actual solutions,
as illustrated in Fig.\ \ref{fig.RK4}.

\begin{figure}[htp]
\includegraphics[scale=0.5]{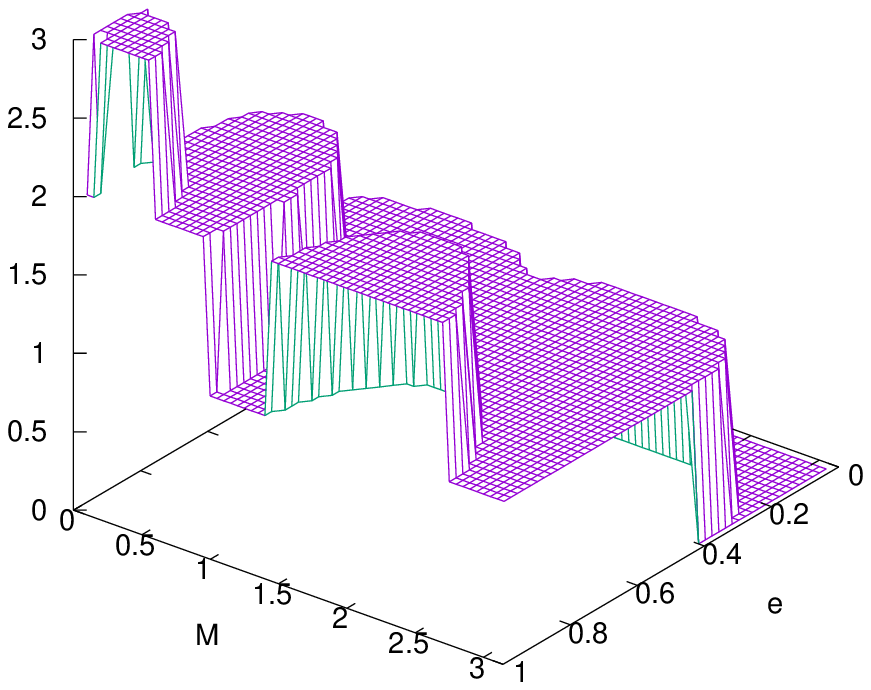}
\caption{
The number of iterations needed for a relative accuracy of $10^{-12}$ in $E$
starting from the RK4 solution of \eqref{eq.RK4} iterating with (\ref{eq.N1}).
}
\label{fig.dgsRK41}
\end{figure}

Fig.\ \ref{fig.dgsRK41} demonstrates that 3 steps of the first-order Newton 
method suffice to obtain 12 digits  accuracy in $E$.

This approach is close in spirit to the Taylor series \eqref{eq.Eofe},
but not suffering from the singularity at $e\to 1$.

\begin{figure}[htp]
\includegraphics[scale=0.5]{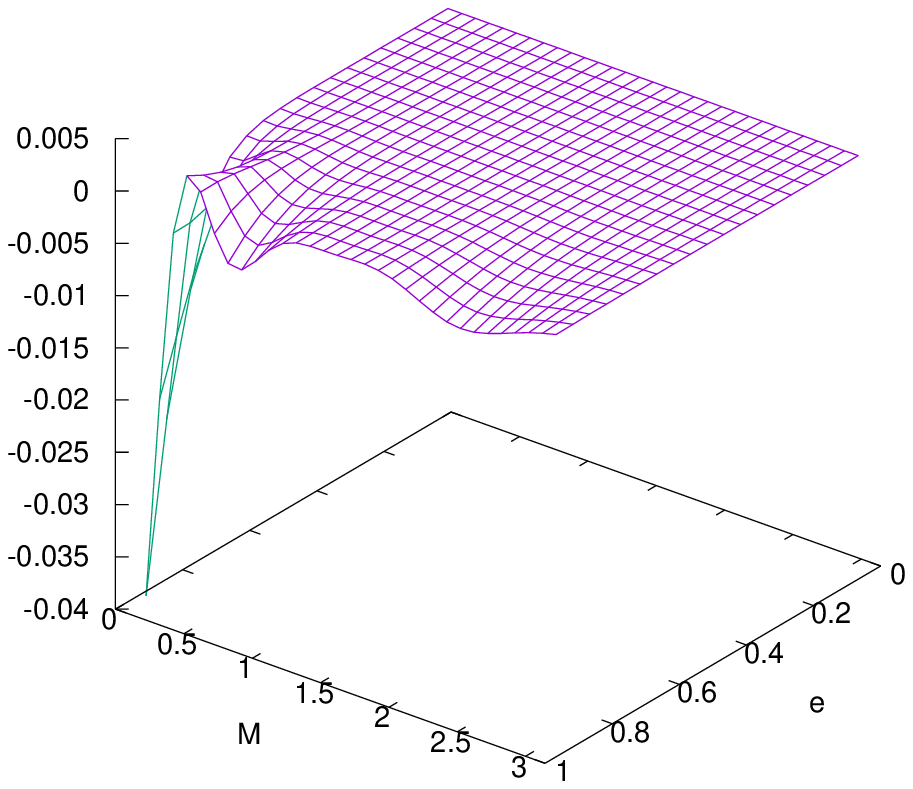}
\caption{Mismatch $E^{(0)}-E$ of the initial estimate of a single-step
5th-order Runge-Kutta-Fehlberg estimator for \eqref{eq.RK4}.
}
\label{fig.RK5}
\end{figure}

\begin{figure}[htp]
\includegraphics[width=0.5\textwidth]{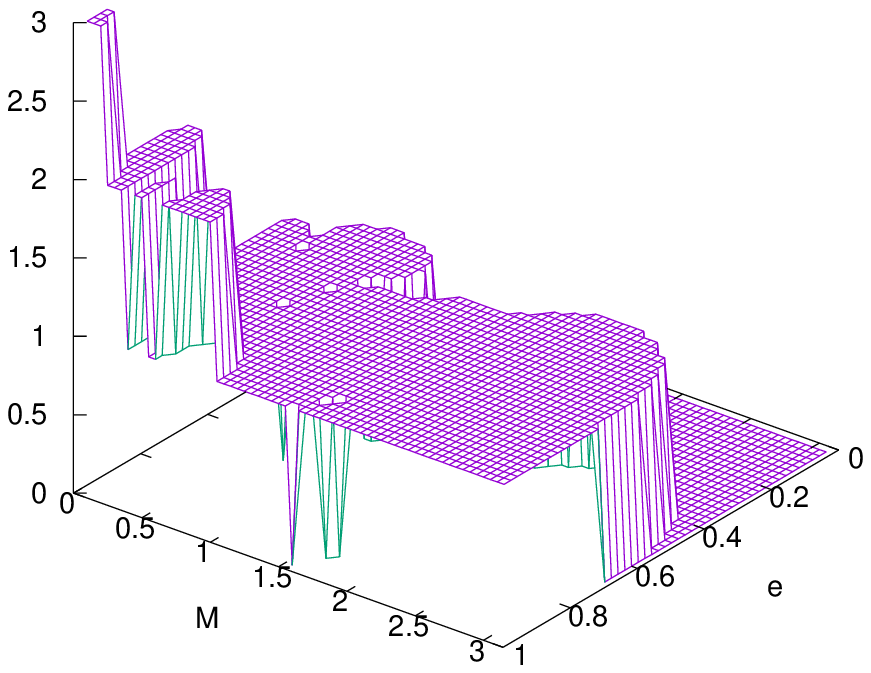}
\caption{
The number of iterations needed for a relative accuracy of $10^{-12}$ in $E$
starting from the RK5(6) solution of \eqref{eq.RK4} iterating with (\ref{eq.N1}).
}
\label{fig.dgsRK51}
\end{figure}

For higher accuracy we can also solve this with an ``explicit'' 5-th order Runge-Kutta-Fehlberg
RK5(6) method, \cite[Table III]{FehlbergTRR287}\cite[Tab. 1]{FehlbergComp4}. The error
in these guesses and the number of iterations in the Newton methods are summarized
in Figures \ref{fig.RK5} and \ref{fig.dgsRK51}.

\section{Summary}
A initial value (\ref{eq.E0qu}) combined with Halley's equation (\ref{eq.N2})
leads to fast and stable
convergence for the inverse problem of Kepler's equation for elliptic orbits.
That initial value is simpler but generally \emph{worse} than Markley's estimator \cite{MarkleyCM63}.

\clearpage

\appendix
\section{C++ Implementation}
A reference implementation is reprinted in the \texttt{anc} directory which implements
(\ref{eq.N2}) and (\ref{eq.E0qu}).
If compiled with the \texttt{-DTEST} preprocessor symbol,
\texttt{g++ -O2 -DTEST -o solveKepler solveKepler.cc},
a test program is obtained which can be called with
one option and two command line arguments, \texttt{solveKepler } \textit{[-1|-2|-3]} $e$ $M$, to investigate
the convergence with this and other approaches.
The option \texttt{-1} indicates that iteration with \eqref{eq.N1} computes
successive approximations, the option {-2} (or none) indicates iteration
with \eqref{eq.N2} and the option {-3} iterations with Householder's
method including third derivatives $f'''$ \cite{AbbasbandyAMC145}.
It prints for all implemented starting guesses
the values of $e$, $M$, the index $i$
of the iteration loop in the Newton method, the estimate $E^{(i)}$
reached so far, and the error $E^{(i)}-E$ relative to the true solution.

The program switches over to solving
\begin{equation}
M = e \sinh E -E
\end{equation}
with the estimate $E^{(0)}= \arsinh(M/e)$
for hyperbolic orbits if the command line parameter $e$ is larger than 1.

\bibliography{all}

\begin{thebibliography}{27}%
\makeatletter
\providecommand \@ifxundefined [1]{%
 \@ifx{#1\undefined}
}%
\providecommand \@ifnum [1]{%
 \ifnum #1\expandafter \@firstoftwo
 \else \expandafter \@secondoftwo
 \fi
}%
\providecommand \@ifx [1]{%
 \ifx #1\expandafter \@firstoftwo
 \else \expandafter \@secondoftwo
 \fi
}%
\providecommand \natexlab [1]{#1}%
\providecommand \enquote  [1]{``#1''}%
\providecommand \bibnamefont  [1]{#1}%
\providecommand \bibfnamefont [1]{#1}%
\providecommand \citenamefont [1]{#1}%
\providecommand \href@noop [0]{\@secondoftwo}%
\providecommand \href [0]{\begingroup \@sanitize@url \@href}%
\providecommand \@href[1]{\@@startlink{#1}\@@href}%
\providecommand \@@href[1]{\endgroup#1\@@endlink}%
\providecommand \@sanitize@url [0]{\catcode `\\12\catcode `\$12\catcode
  `\&12\catcode `\#12\catcode `\^12\catcode `\_12\catcode `\%12\relax}%
\providecommand \@@startlink[1]{}%
\providecommand \@@endlink[0]{}%
\providecommand \url  [0]{\begingroup\@sanitize@url \@url }%
\providecommand \@url [1]{\endgroup\@href {#1}{\urlprefix }}%
\providecommand \urlprefix  [0]{URL }%
\providecommand \Eprint [0]{\href }%
\providecommand \doibase [0]{https://doi.org/}%
\providecommand \selectlanguage [0]{\@gobble}%
\providecommand \bibinfo  [0]{\@secondoftwo}%
\providecommand \bibfield  [0]{\@secondoftwo}%
\providecommand \translation [1]{[#1]}%
\providecommand \BibitemOpen [0]{}%
\providecommand \bibitemStop [0]{}%
\providecommand \bibitemNoStop [0]{.\EOS\space}%
\providecommand \EOS [0]{\spacefactor3000\relax}%
\providecommand \BibitemShut  [1]{\csname bibitem#1\endcsname}%
\let\auto@bib@innerbib\@empty
\bibitem [{\citenamefont {Fitzpatrick}(1970)}]{FitzpatrickPM}%
  \BibitemOpen
  \bibfield  {author} {\bibinfo {author} {\bibfnamefont {P.~M.}\ \bibnamefont
  {Fitzpatrick}},\ }\href@noop {} {\emph {\bibinfo {title} {Principles of
  celestial mechanics}}}\ (\bibinfo  {publisher} {Academic Press},\ \bibinfo
  {year} {1970})\BibitemShut {NoStop}%
\bibitem [{\citenamefont {Danby}\ and\ \citenamefont
  {Burkardt}(1983)}]{DanbyCM31}%
  \BibitemOpen
  \bibfield  {author} {\bibinfo {author} {\bibfnamefont {J.~M.~A.}\
  \bibnamefont {Danby}}\ and\ \bibinfo {author} {\bibfnamefont {T.~M.}\
  \bibnamefont {Burkardt}},\ }\bibfield  {title} {\bibinfo {title} {The
  solution of kepler's equation. i},\ }\href
  {https://doi.org/10.1007/BF01686811} {\bibfield  {journal} {\bibinfo
  {journal} {Cel. Mech.}\ }\textbf {\bibinfo {volume} {31}},\ \bibinfo {pages}
  {95} (\bibinfo {year} {1983})}\BibitemShut {NoStop}%
\bibitem [{\citenamefont {Gerlach}(1994)}]{GerlachSIAM36}%
  \BibitemOpen
  \bibfield  {author} {\bibinfo {author} {\bibfnamefont {J.}~\bibnamefont
  {Gerlach}},\ }\bibfield  {title} {\bibinfo {title} {Accelerated convergence
  in {N}ewton's method},\ }\href {https://doi.org/10.1137/1036057} {\bibfield
  {journal} {\bibinfo  {journal} {SIAM Review}\ }\textbf {\bibinfo {volume}
  {36}},\ \bibinfo {pages} {272} (\bibinfo {year} {1994})}\BibitemShut
  {NoStop}%
\bibitem [{\citenamefont {Hansen}\ and\ \citenamefont
  {Patrick}(1977)}]{HansenNumerMath27}%
  \BibitemOpen
  \bibfield  {author} {\bibinfo {author} {\bibfnamefont {E.}~\bibnamefont
  {Hansen}}\ and\ \bibinfo {author} {\bibfnamefont {M.}~\bibnamefont
  {Patrick}},\ }\bibfield  {title} {\bibinfo {title} {A family of root finding
  methods},\ }\href {https://doi.org/10.1007/BF01396176} {\bibfield  {journal}
  {\bibinfo  {journal} {Numer.\ Math.}\ }\textbf {\bibinfo {volume} {27}},\
  \bibinfo {pages} {257} (\bibinfo {year} {1977})}\BibitemShut {NoStop}%
\bibitem [{\citenamefont {Kalantari}\ \emph {et~al.}(1997)\citenamefont
  {Kalantari}, \citenamefont {Kalantari},\ and\ \citenamefont
  {Zaare-Nahandi}}]{KalantariJCAM80}%
  \BibitemOpen
  \bibfield  {author} {\bibinfo {author} {\bibfnamefont {B.}~\bibnamefont
  {Kalantari}}, \bibinfo {author} {\bibfnamefont {I.}~\bibnamefont
  {Kalantari}},\ and\ \bibinfo {author} {\bibfnamefont {R.}~\bibnamefont
  {Zaare-Nahandi}},\ }\bibfield  {title} {\bibinfo {title} {A basic family of
  iteration functions for polynomial root finding and its characterizations},\
  }\href {https://doi.org/10.1016/S0377-0427(97)00014-9} {\bibfield  {journal}
  {\bibinfo  {journal} {J. Comp.\ Appl.\ Math.}\ }\textbf {\bibinfo {volume}
  {80}},\ \bibinfo {pages} {209} (\bibinfo {year} {1997})}\BibitemShut
  {NoStop}%
\bibitem [{\citenamefont {Alefeld}(1981)}]{AlefeldAMM88}%
  \BibitemOpen
  \bibfield  {author} {\bibinfo {author} {\bibfnamefont {G.}~\bibnamefont
  {Alefeld}},\ }\bibfield  {title} {\bibinfo {title} {On the convergence of
  halley's method},\ }\href {https://doi.org/10.2307/2321760} {\bibfield
  {journal} {\bibinfo  {journal} {Am. Math. Monthly}\ }\textbf {\bibinfo
  {volume} {88}},\ \bibinfo {pages} {530} (\bibinfo {year} {1981})}\BibitemShut
  {NoStop}%
\bibitem [{\citenamefont {Esmaelzadeh}\ and\ \citenamefont
  {Ghadiri}(2014)}]{EsmaelIJCA89}%
  \BibitemOpen
  \bibfield  {author} {\bibinfo {author} {\bibfnamefont {R.}~\bibnamefont
  {Esmaelzadeh}}\ and\ \bibinfo {author} {\bibfnamefont {H.}~\bibnamefont
  {Ghadiri}},\ }\bibfield  {title} {\bibinfo {title} {Appropriate starter for
  solving the kepler's equation},\ }\href {https://doi.org/10.5120/15517-4394}
  {\bibfield  {journal} {\bibinfo  {journal} {Int. J. Comp. Applic.}\ }\textbf
  {\bibinfo {volume} {89}},\ \bibinfo {pages} {31} (\bibinfo {year}
  {2014})}\BibitemShut {NoStop}%
\bibitem [{\citenamefont {{International Astronomical Union}}(2023)}]{SOFA}%
  \BibitemOpen
  \bibfield  {author} {\bibinfo {author} {\bibnamefont {{International
  Astronomical Union}}},\ }\href {http://www.iausofa.org} {\emph {\bibinfo
  {title} {IAU SOFA Collection}}},\ \bibinfo {type} {Tech. Rep.}\ (\bibinfo
  {institution} {International Astronomical Union},\ \bibinfo {year}
  {2023})\BibitemShut {NoStop}%
\bibitem [{\citenamefont {Fukushima}(1997)}]{FukushimaCM66}%
  \BibitemOpen
  \bibfield  {author} {\bibinfo {author} {\bibfnamefont {T.}~\bibnamefont
  {Fukushima}},\ }\bibfield  {title} {\bibinfo {title} {A method solving
  kepler's equation without transcendental function evaluations},\ }\href
  {https://doi.org/10.1007/BF00049384} {\bibfield  {journal} {\bibinfo
  {journal} {Cel. Mech. Dyn. Astr.}\ }\textbf {\bibinfo {volume} {66}},\
  \bibinfo {pages} {309} (\bibinfo {year} {1997})}\BibitemShut {NoStop}%
\bibitem [{\citenamefont {Palacios}(2002)}]{PalaciosJCAM138}%
  \BibitemOpen
  \bibfield  {author} {\bibinfo {author} {\bibfnamefont {M.}~\bibnamefont
  {Palacios}},\ }\bibfield  {title} {\bibinfo {title} {Kepler equation and
  accelerated newton method},\ }\href
  {https://doi.org/10.1016/S0377-0427(01)00369-7} {\bibfield  {journal}
  {\bibinfo  {journal} {J. Comp. Appl. Math}\ }\textbf {\bibinfo {volume}
  {138}},\ \bibinfo {pages} {335} (\bibinfo {year} {2002})}\BibitemShut
  {NoStop}%
\bibitem [{\citenamefont {Grau-S\'anchez}\ \emph {et~al.}(2011)\citenamefont
  {Grau-S\'anchez}, \citenamefont {Grau},\ and\ \citenamefont
  {Noguera}}]{GrauAMC218}%
  \BibitemOpen
  \bibfield  {author} {\bibinfo {author} {\bibfnamefont {M.}~\bibnamefont
  {Grau-S\'anchez}}, \bibinfo {author} {\bibfnamefont {A.}~\bibnamefont
  {Grau}},\ and\ \bibinfo {author} {\bibfnamefont {M.}~\bibnamefont
  {Noguera}},\ }\bibfield  {title} {\bibinfo {title} {Ostrowki type methods for
  solving systems of nonlinear equations},\ }\href
  {https://doi.org/10.1016/j.amc.2011.08.011} {\bibfield  {journal} {\bibinfo
  {journal} {Appl. Math. Comput.}\ }\textbf {\bibinfo {volume} {218}},\
  \bibinfo {pages} {2377} (\bibinfo {year} {2011})}\BibitemShut {NoStop}%
\bibitem [{\citenamefont {Chun}(2007)}]{ChunAMC190}%
  \BibitemOpen
  \bibfield  {author} {\bibinfo {author} {\bibfnamefont {C.}~\bibnamefont
  {Chun}},\ }\bibfield  {title} {\bibinfo {title} {Some variants of king's
  fourth-order family of methods for nonlinear equations},\ }\href
  {https://doi.org/10.1016/j.amc.2007.01.006} {\bibfield  {journal} {\bibinfo
  {journal} {Appl. Math. Comp.}\ }\textbf {\bibinfo {volume} {190}},\ \bibinfo
  {pages} {57} (\bibinfo {year} {2007})}\BibitemShut {NoStop}%
\bibitem [{\citenamefont {Conway}(1986)}]{ConwayAIAA86}%
  \BibitemOpen
  \bibfield  {author} {\bibinfo {author} {\bibfnamefont {B.~A.}\ \bibnamefont
  {Conway}},\ }\bibfield  {title} {\bibinfo {title} {An improved algorithm due
  to laguerre for the solution of kepler's equation},\ }in\ \href
  {https://doi.org/10.2514/6.1986-84} {\emph {\bibinfo {booktitle} {24th
  Aerospace Sciences Meeting}}},\ \bibinfo {series and number} {\bibinfo
  {number} {86-0083}}\ (\bibinfo  {publisher} {AIAA},\ \bibinfo {year}
  {1986})\BibitemShut {NoStop}%
\bibitem [{\citenamefont {Charles}\ and\ \citenamefont
  {Tatum}(1997)}]{CharlesCMDA69}%
  \BibitemOpen
  \bibfield  {author} {\bibinfo {author} {\bibfnamefont {E.~D.}\ \bibnamefont
  {Charles}}\ and\ \bibinfo {author} {\bibfnamefont {J.~B.}\ \bibnamefont
  {Tatum}},\ }\bibfield  {title} {\bibinfo {title} {The convergence of
  newton-raphson iteration with kepler's equation},\ }\href
  {https://doi.org/10.1023/A:1008200607490} {\bibfield  {journal} {\bibinfo
  {journal} {Cel. Mech. Dyn. Astr.}\ }\textbf {\bibinfo {volume} {69}},\
  \bibinfo {pages} {357} (\bibinfo {year} {1997})}\BibitemShut {NoStop}%
\bibitem [{\citenamefont {Stumpf}(1999)}]{StumpfCM74}%
  \BibitemOpen
  \bibfield  {author} {\bibinfo {author} {\bibfnamefont {L.}~\bibnamefont
  {Stumpf}},\ }\bibfield  {title} {\bibinfo {title} {Chaotic behaviour in the
  newton iterative function associated with kepler's equation},\ }\href
  {https://doi.org/10.1023/A:1008339416143} {\bibfield  {journal} {\bibinfo
  {journal} {Cel. Mech. Dyn. Astr.}\ }\textbf {\bibinfo {volume} {74}},\
  \bibinfo {pages} {95} (\bibinfo {year} {1999})}\BibitemShut {NoStop}%
\bibitem [{\citenamefont {Abramowitz}\ and\ \citenamefont {Stegun}(1972)}]{AS}%
  \BibitemOpen
  \bibinfo {editor} {\bibfnamefont {M.}~\bibnamefont {Abramowitz}}\ and\
  \bibinfo {editor} {\bibfnamefont {I.~A.}\ \bibnamefont {Stegun}},\ eds.,\
  \href {https://archive.org/details/HandbookOfMathematicalFunctions} {\emph
  {\bibinfo {title} {Handbook of Mathematical Functions}}},\ \bibinfo {edition}
  {9th}\ ed.\ (\bibinfo  {publisher} {Dover Publications},\ \bibinfo {address}
  {New York},\ \bibinfo {year} {1972})\BibitemShut {NoStop}%
\bibitem [{\citenamefont {Schonfelder}(1980)}]{Schonfelder}%
  \BibitemOpen
  \bibfield  {author} {\bibinfo {author} {\bibfnamefont {J.~L.}\ \bibnamefont
  {Schonfelder}},\ }\bibfield  {title} {\bibinfo {title} {Very high accuracy
  {C}hebyshev expansions for the basic trigonometric functions},\ }\href
  {https://doi.org/10.1090/S0025-5718-1980-0551302-5} {\bibfield  {journal}
  {\bibinfo  {journal} {Math.\ Comp.}\ }\textbf {\bibinfo {volume} {34}},\
  \bibinfo {pages} {237} (\bibinfo {year} {1980})}\BibitemShut {NoStop}%
\bibitem [{\citenamefont {Boyd}(2007)}]{BoydANM57}%
  \BibitemOpen
  \bibfield  {author} {\bibinfo {author} {\bibfnamefont {J.~P.}\ \bibnamefont
  {Boyd}},\ }\bibfield  {title} {\bibinfo {title} {Rootfinding for a
  transcendental equation without a first guess: polynomialization of kepler's
  equation through chebyshev polyomial expansion of the sine},\ }\href
  {https://doi.org/10.1016/j.apnum.2005.11.010} {\bibfield  {journal} {\bibinfo
   {journal} {Appl. Num. Math.}\ }\textbf {\bibinfo {volume} {57}},\ \bibinfo
  {pages} {12} (\bibinfo {year} {2007})}\BibitemShut {NoStop}%
\bibitem [{\citenamefont {Boyd}(2009)}]{BoydJCAM223}%
  \BibitemOpen
  \bibfield  {author} {\bibinfo {author} {\bibfnamefont {J.~P.}\ \bibnamefont
  {Boyd}},\ }\bibfield  {title} {\bibinfo {title} {Chebyshev expansion in
  intervals with branch points with application to the root of kepler's
  equation: A chebyshev-hermite-pad\'e method},\ }\href
  {https://doi.org/10.1016/j.cam.2008.02.007} {\bibfield  {journal} {\bibinfo
  {journal} {J. Comp. Appl. Math.}\ }\textbf {\bibinfo {volume} {223}},\
  \bibinfo {pages} {693} (\bibinfo {year} {2009})}\BibitemShut {NoStop}%
\bibitem [{\citenamefont {Mikkola}(1987)}]{MikkolaCM40}%
  \BibitemOpen
  \bibfield  {author} {\bibinfo {author} {\bibfnamefont {S.}~\bibnamefont
  {Mikkola}},\ }\bibfield  {title} {\bibinfo {title} {A cubic approximation for
  kepler's equation},\ }\href {https://doi.org/10.1007/BF01235850} {\bibfield
  {journal} {\bibinfo  {journal} {Cel.\ Mech.}\ }\textbf {\bibinfo {volume}
  {40}},\ \bibinfo {pages} {329} (\bibinfo {year} {1987})}\BibitemShut
  {NoStop}%
\bibitem [{\citenamefont {Nijenhuis}(1991)}]{NijenhuisCM51}%
  \BibitemOpen
  \bibfield  {author} {\bibinfo {author} {\bibfnamefont {A.}~\bibnamefont
  {Nijenhuis}},\ }\bibfield  {title} {\bibinfo {title} {Solving kepler's
  equation with high efficiency and accuracy},\ }\href
  {https://doi.org/10.1007/BF00052925} {\bibfield  {journal} {\bibinfo
  {journal} {Cel. Mech. Dyn. Astr.}\ }\textbf {\bibinfo {volume} {51}},\
  \bibinfo {pages} {319} (\bibinfo {year} {1991})}\BibitemShut {NoStop}%
\bibitem [{\citenamefont {Markley}(1995)}]{MarkleyCM63}%
  \BibitemOpen
  \bibfield  {author} {\bibinfo {author} {\bibfnamefont {F.~L.}\ \bibnamefont
  {Markley}},\ }\bibfield  {title} {\bibinfo {title} {Kepler equation solver},\
  }\href {https://doi.org/10.1007/BF00691917} {\bibfield  {journal} {\bibinfo
  {journal} {Cel. Mech. Dyn. Astr.}\ }\textbf {\bibinfo {volume} {63}},\
  \bibinfo {pages} {101} (\bibinfo {year} {1995})}\BibitemShut {NoStop}%
\bibitem [{\citenamefont {Zurm\"uhl}(1948)}]{ZurmuhlZAMM28}%
  \BibitemOpen
  \bibfield  {author} {\bibinfo {author} {\bibfnamefont {R.}~\bibnamefont
  {Zurm\"uhl}},\ }\bibfield  {title} {\bibinfo {title} {Runge-kutta-verfahren
  zur numerischen {I}ntegration von {D}ifferentialgleichungen n-ter
  {O}rdnung},\ }\href {https://doi.org/10.1002/zamm.19480280603} {\bibfield
  {journal} {\bibinfo  {journal} {Z. angew. Math. Mech.}\ }\textbf {\bibinfo
  {volume} {28}},\ \bibinfo {pages} {173} (\bibinfo {year} {1948})}\BibitemShut
  {NoStop}%
\bibitem [{\citenamefont {Butcher}(1963)}]{ButcherJAMS3}%
  \BibitemOpen
  \bibfield  {author} {\bibinfo {author} {\bibfnamefont {J.~C.}\ \bibnamefont
  {Butcher}},\ }\bibfield  {title} {\bibinfo {title} {Coefficients for the
  study of runge-kutta integration processes},\ }\href
  {https://doi.org/10.1017/S1446788700027932} {\bibfield  {journal} {\bibinfo
  {journal} {J. Austral. Math. Soc.}\ }\textbf {\bibinfo {volume} {3}},\
  \bibinfo {pages} {185} (\bibinfo {year} {1963})}\BibitemShut {NoStop}%
\bibitem [{\citenamefont {Fehlberg}(1968)}]{FehlbergTRR287}%
  \BibitemOpen
  \bibfield  {author} {\bibinfo {author} {\bibfnamefont {E.}~\bibnamefont
  {Fehlberg}},\ }\bibfield  {title} {\bibinfo {title} {Classical fifth-,
  sixth-, seventh-, and eigth-order runge-kutte formulas with stepsize
  control},\ }\href {https://ntrs.nasa.gov/citations/19680027281} {\  (\bibinfo
  {year} {1968})}\BibitemShut {NoStop}%
\bibitem [{\citenamefont {Fehlberg}(1969)}]{FehlbergComp4}%
  \BibitemOpen
  \bibfield  {author} {\bibinfo {author} {\bibfnamefont {E.}~\bibnamefont
  {Fehlberg}},\ }\bibfield  {title} {\bibinfo {title} {Klassische
  runge-kutte-formeln f\"unfter und siebenter ordnung mit
  schrittweiten-kontrolle},\ }\href {https://doi.org/10.1007/BF02234758}
  {\bibfield  {journal} {\bibinfo  {journal} {Computing}\ }\textbf {\bibinfo
  {volume} {4}},\ \bibinfo {pages} {93} (\bibinfo {year} {1969})}\BibitemShut
  {NoStop}%
\bibitem [{\citenamefont {Abbasbandy}(2003)}]{AbbasbandyAMC145}%
  \BibitemOpen
  \bibfield  {author} {\bibinfo {author} {\bibfnamefont {S.}~\bibnamefont
  {Abbasbandy}},\ }\bibfield  {title} {\bibinfo {title} {Improving
  newton-raphson method for nonlinear equations by modified adomian
  decomposition method},\ }\href
  {https://doi.org/10.1016/S0096-3003(03)00282-0} {\bibfield  {journal}
  {\bibinfo  {journal} {Appl. Math. Comput.}\ }\textbf {\bibinfo {volume}
  {145}},\ \bibinfo {pages} {887} (\bibinfo {year} {2003})}\BibitemShut
  {NoStop}%
\end{thebibliography}%

\end{document}